\input harvmac
\input epsf
%\draftmode

\lref\gvconjecture{R. Gopakumar and C. Vafa, 
``On the Gauge Theory/Geometry Correspondence,''
{\tt hep-th/9811131}.}
\lref\fp{C. Faber and R. Pandharipande,
``Hodge Integrals and Gromov-Witten Theory,'':
{\tt math.AG/9810173}.}
\lref\candelas{P. Candelas, X.C. De La Ossa, P. S. Green and
L. Parkes, ``A Pair Of Calabi-Yau Manifolds As An Exactly Soluble 
Superconformal Theory,'' Nucl. Phys. B359 (1991) 21.}
\lref\bcovone{M. Bershadsky, S. Cecotti,
H. Ooguri and C. Vafa, ``Holomorphic Anomalies
in Topological Field Theories,'' Nucl. Phys. B405 (1993) 279;
{\tt hep-th/9302103}.}
\lref\bcovtwo{M. Bershadsky, S. Cecotti,
H. Ooguri and C. Vafa, ``Kodaira-Spencer Theory
of Gravity and Exact Results for Quantum 
String Theory,'' Commun. Math. Phys. 165 (1994) 311; 
{\tt hep-th/9309140}.}
\lref\wittenopen{E. Witten, 
`` Chern-Simons Gauge Theory As A String Theory,''
{\tt hep-th/9207094}.}
\lref\wittencs{E. Witten,
``Quantum Field Theory And The Jones Polynomial,''
 Commun. Math. Phys. 121 (1989) 351.}

\noblackbox

\Title{\vbox{\baselineskip12pt
\hbox{\tt HUTP-00/A050}
\hbox{\tt hep-th/0012136}}}%
{\vbox{\centerline{SO and Sp Chern-Simons at Large N}
}}
\bigskip
\centerline{Shishir Sinha and Cumrun Vafa}
\vglue 1cm
\smallskip
\centerline{ Jefferson Physical Laboratory}
\centerline{
Harvard University}
\centerline{ Cambridge, MA 02138, USA}

\vskip 0.5cm
\vskip .3in
We study the large $N$ limit of $SO(N)$ and $Sp(N)$
Chern-Simons gauge theory on $S^3$ and identify its
closed string dual as topological strings on an 
orientifold of the small resolution of the conifold.
Applications to large $N$ dualities for
 ${\cal N}=1$ supersymmetric gauge systems in 
4 dimensions are also discussed.
\Date{December 2000}

\newsec{Introduction}

The aim of this note is to extend the large $N$ duality conjecture
of \gvconjecture\ which relates $SU(N)$ Chern-Simons
gauge theory on $S^3$ to topological strings on the small resolution
of the conifold, to the case of the $SO(N)$ and $Sp(N)$ gauge theories.
The basic idea is to act by orientifolding on the duality of \gvconjecture\
and obtain the new duality for $SO(N)$ and $Sp(N)$ cases.
The main subtlety arises in the exact identification of the parameters
on both sides.  This we fix by various consistency checks.

The conjecture of \gvconjecture\ has been embedded in type II
superstring theory
in \ref\vaau{C. Vafa, ``Superstrings and Topological 
Strings at Large N,'' hep-th/0008142.}. Furthermore this duality
has been restated in purely geometric setup by embedding type
IIa superstring in M-theory \ref\amv{M. Atiyah,
J. Maldacena, and C. Vafa, ``An M-theory Flop as a Large 
N Duality,'' hep-th/0011256.}.
Similarly we can raise the same questions for the case of $SO$ and
$Sp$ gauge group.  In particular we show how to embed this duality 
in type IIA superstrings and interpret it in purely geometric
terms by further embedding it in M-theory.

The organization of this paper is as follows:  In section
2 we review the Large $N$ conjecture for $SU(N)$ Chern-Simons
theory.  In section 3 we propose a large $N$ conjecture
for $SO(N)$ and $Sp(N)$ Chern-Simons theory on $S^3$.
In section 4 we consider the partition function of the
Chern-Simons theory for these classes of gauge groups
and perform a large $N$ expansion.  In section 5 we compare
the results with expectations based on the conjectured
large $N$ dual.  In section 6 we consider connections with
${\cal N}=1$ systems in 4 dimensions.

\newsec{The Large $N$ conjecture for $SU(N)$ Chern-Simons Theory }

In this section we briefly review the conjecture of \gvconjecture\ 
which relates large $N$ limit of $SU(N)$ Chern-Simons
gauge theory on $S^3$ to a particular topological string.
 The conjecture in \gvconjecture\ 
states that
the Chern-Simons gauge theory on $S^3$ with gauge
group $SU(N)$ and level $k$ is equivalent, at large $N$, to
the closed topological string theory of $A$-type
on the $S^2$ blown up conifold geometry with 
\eqn\parameters{g_s = {2\pi i \over k+N}, ~~~~ 
t={2\pi i N \over k+N},}
where $g_s$ is the string coupling constant
and $t$ is the K\"ahler modulus of the blown-up $S^2$. 
The coupling constant $g_{CS}$ of the Chern-Simons theory, 
after taking into account the finite renormalization, is
related to $g_s$ as $g_s=g_{CS}^2$. Therefore the K\"ahler moduli $t$
given by \parameters\ is 
the 't Hooft coupling $g_{CS}^2 N$ of the 
Chern-Simons theory.

 The geometric motivation of the
conjecture is based on starting with the topological strings on conifold
geometry $T^*S^3$ and
putting many branes on $S^3$, for which we get a large $N$
limit of Chern-Simons on $S^3$ supported on the brane \wittenopen .
The conjecture states that in the large
$N$ limit the branes disappear but the geometry gets deformed and
 an $S^2$ gets
blown up.

The conjecture has been checked for the free energy
to all orders in the
$1/N$ expansion (since both sides are computable) as well
as a large class of Wilson loop expectation values 
\ref\ov{H. Ooguri and C. Vafa,``Knot Invariants and Topological Strings,''
Nucl. Phys. {\bf B577} (2000) 419.}\ref\labm{
J.M.F. Labastida and M. Marino,``Polynomial invariants 
for torus knots and topological strings,'' hep-th/0004196.}\ref\sar{P.
Ramadevi and T. Sarkar,``On Link Invariants and Topological
 String Amplitudes,'' hep-th/0009188.}\ref\lmv{
J.M.F. Labastida and M. Marino and C. Vafa,``Knots, 
links and branes at large N,'' JHEP 0011 (2000) 007.}.
  In particular the partition function
$Z_{SU}(S^3)$ of the $SU(N)$
Chern-Simons gauge theory on $S^3$ is given by
\eqn\cspartition{Z_{SU}(S^3) = {e^{i{\pi \over 2}(N-1)N} \over (k+N)^{N/2}}
\sqrt{{k+N \over n}} \prod_{s=1}^{N-1} \left[2 \sin\left({N\over k+N}
\right) \right]^{N-s}.}
The large-$N$ expansion of $\log Z_{SU}(S^3)$ is given by
\eqn\largenpartition{
Z_{SU}(S^3) = \exp\left[  \sum_{g=0}^\infty g_s^{2g-2} F_g(t) \right],}
where $g_s$ and $t$ as in \parameters ,
\eqn\zeroandone{
\eqalign{ F_0 & = -\zeta(3) + {i\pi^2\over 6} t -i\left( m + {1\over 4}\right)
\pi t^2 + {i \over 12}t^3 + \sum_{n=1}^\infty n^{-3} e^{-nt} \cr
F_1 & = {1 \over 24} t + {1 \over 12} \log \left( 1-e^{-t} \right) ,\cr}}
with $m$ being some integer, and for $g \geq 2$, 
\eqn\twoorlarger{
F_g = {(-1)^{g-1} \over 2g(2g-2)} B_g \left[ 
{(-1)^{g-1} \over (2\pi)^{2g-2}} 2\zeta(2g-2)
        - {1 \over (2g-3)!} \sum_{n=1}^\infty n^{2g-3} e^{-nt}
\right].}
Here $B_g$ is the Bernoulli number.
It turns out that the expressions \zeroandone\ and \twoorlarger\
for $F_g$ are exactly those of the $g$-loop topological
string amplitude on the resolved conifold. 
These expressions can be derived, as was done in \ref\gopva{
R. Gopakumar and C. Vafa, ``M-Theory and Topological Strings I, II,''
{\tt hep-th/9809197, 9812127}.}, from the target space view point by
identifying what the topological strings compute in Type IIA
compactification on the corresponding Calabi-Yau space. 
They can also be derived using the mathematical
definition of topological string
amplitudes \fp .

The geometric transition underlying this large $N$ duality is
the conifold transition which is reviewed below.

\subsec{Conifold transition}

 Conifold is described by $z_1z_4 - z_2z_3 = \mu$ . It is a non-compact
 Calabi Yau manifold. This manifold is also described by $T^*S^3$. The
 topology of the manifold is that of a deformed cone with base = $S^2
 \times S^3$. To see this by substituting
\eqn\sub{ \eqalign{ z_1 = y_1+iy_2 \cr
          z_2 = -y_3+iy_4 \cr
          z_3 = y_3+iy_4 \cr
          z_4 = y_1-iy_2 }}
 the conifold equation becomes $y_1^2 + y_2^2 +y_3^2 +y_4^2 = \mu$. Taking
 the real section of this we notice that there is an $S^3$
 (and the imaginary parts give the cotangent directions). As $\mu = 0$, 
$S^3$ shrinks to zero size and the manifold becomes singular. 
The singularity can be removed by what is called
 a ``small resolution'' which is to replace the origin $z_i=0$ by
 an $S^2$.  In particular the $S^2$ is parameterized by
 a complex coordinate $z$ which is defined by
$z_1 = z z_2$  or $z_3 = z z_4$  (note that the two
are consistent because $z_1z_4-z_2z_3=0$).  The resulting
manifold (in some patch) is now 
 described by three complex coordinates $(z, z_1,
z_4)$.  This is what is called
the conifold transition.  We go from having an
$S^3$ submanifold at the tip of the cone to having an $S^2$.
The topological string for $SU(N)$ Chern-Simons gauge theory is obtained
by having $N$ D3 branes wrapping the $S^3$.  At large $N$ the better
description is in terms of the blowup geometry where the $S^3$
is replaced by $S^2$ of finite size and the branes have disappeared.

\newsec{The $SO(N)$ and $Sp(N)$ Chern-Simons Duals}

In this section we extend the duality of \gvconjecture\ to the
case of $SO$ and $Sp$ gauge groups.  The basic idea is to start
with the duality of \gvconjecture\ for the $SU(N)$ gauge group
and orientifold both sides.  On the side of the
conifold with finite size $S^3$ the orientifolding
should fix the $S^3$.  This leads, depending on the choice
of the sign for worldsheets with crosscaps, to an $SO(N)$ or $Sp(N)$ Chern-Simons
gauge theory, as is familiar in the context of D-branes.

 We consider the following involution
  in the $S^3$ conifold geometry:
\eqn\ident{ z_{1}\rightarrow \bar {z_{4}};~~~~~~~ z_{2}\rightarrow -\bar {z_{3}}}
Note that this $Z_2$ operation leaves the $S^3$ invariant.
  Orientifolding the conifold geometry by this involution and
placing D-branes on $S^3$ leads to $SO(N)$ or $Sp(N)$ Chern-Simons
gauge theories on $S^3$.

To find out what the dual of this theory is, all we have to do
is to orientifold the dual for the $SU(N)$. In other
words we have to orientifold the conifold after the
transition where we have a blown up $S^2$.
To see how this involution acts on the blownup geometry
consider the coordinate chart given by
 $(z_{1}, z_{4}, z)$ . 
From the identification of $z=z_1/z_2$, we see that the involution
\ident\ maps to
\eqn\invo{(z_1,z_4,z)\rightarrow 
(\bar{z_{4}}, \bar{z_{1}}, {-1 \over \bar{z}}).}
Thus we obtain the dual topological strings by orientifolding
the $S^2$ blown up geometry by this involution.
Note  that the orientifolding action takes $z$ which is the
coordinate describing the blown up $S^2$ to $-1 \over \bar{z}$. This 
makes the $S^2$ into $RP^2$.  So now the target space geometry on the
closed string side has an $RP^2$ instead of $S^2$.
 Note also that this orientifolding
has no fixed points (and thus no orientifold planes).
The next thing to do is to find the precise map between parameters
of the two sides which we will now turn to.

\subsec{Large N expansion parameters }

We need to identify the parameters of the gauge theory
with parameters for strings
propagating in  the blow up of the conifold geometry.
This involves identification of the string coupling constant
as well as the size of the $S^2$ with gauge theory parameters.
At the tree level the gauge theory coupling constant, which
should be identified with string coupling is $1/k$.  But just
as in the $SU(N)$ case one expects a shift in $k$.  In particular
from the gauge theory side
$2\pi i\over(k+c_{g})$ is the renormalised coupling.
  Thus it is natural to identify
the string coupling constant, also on the blow up side with that, i.e.
$$g_s ={2\pi i\over k+c_{g}}$$
for $SO(N)$, we have $c_g=N-2$ and for $Sp(N)$ it is
$c_g={N\over 2}+1$ (where in the $Sp$ case $N$ is even, and the rank of
it is $N/2$).  One also has to identify the volume of $S^2$ with
some function of `t Hooft parameter.  In the case of $SU(N)$ the
natural identification was $t=N g_s=Ng_{YM}^2$.  It turns out,
however, that the natural match in the case of $SO$ and $Sp$ groups
is slightly different and we find
$$t=(N +a)g_s.$$
Moreover for $SO$ case $a=-1$ and for $Sp$ case $a=+1$.

To motivate the replacement of $N$ by $N+a$ we proceed
as follows:  The dual string
theory
doesn't
see the number of D branes but it sees the amount of 
D-brane ``flux''.
This notion is not very precise in the case of the topological string
because there is no gauge field coupled to D-brane flux.  Nevertheless
we will use the intuition based on D-branes in ordinary superstrings
and bosonic strings to find the net D-brane ``flux'', which is to
replace $N$ on the dual gravity side.  This comes from the
fact that if we have orientifold planes, they do carry D-brane
flux.  In particular if we have an orientifold plane of dimension
r, in a string theory which has critical dimensions $d$, the D-brane
charge carried by the orientifold plane is
$$a =\mp 2^{{d\over 2}-r}$$
where $\mp$ depends on the choice of the sign
for diagrams with crosscap.  In particular if we have $SO$ 
groups the $-$ sign applies and if we have $Sp$ group the $+$
sign applies.  If we apply this to topological
strings with $d=6$ and noting that the orientifold plane for
us is $S^3$ which has dimension $r=3$, we learn that
$$a=\mp 1.$$
This motivates our choice
of the identification of the size of $S^2$ with gauge theory
parameters.
In the next section we show why these identifications
are also natural from the point of view of gravitational
anomalies for Chern-Simons theory.

\subsec{Anomaly Analysis}

 Since the Chern-Simons theory is a topological theory one expects that it
 is independent of the background metric. The classical lagrangian is
 independent of the background metric. But it was argued \wittencs\
 that at quantum
 level there is a term which does depend on the background metric. It is of
 the form :
 \eqn\anom{ {i\pi c \over 12}\int_{S^3}{(w \wedge dw) + 2/3 
(w \wedge w \wedge w)}}
 Where $w$ is the spin connection and
 $c$ refers to the central charge of the WZW current algebra at level
 $k$ for the group that appears in the Chern-Simons theory. $c$ is given
by $c={k(dim(G))\over{k+c_g}}$. Here $dim(G)$ refers to the number of
 generators that the gauge group has. So for $SO(N)$ case :
 \eqn\val{ c_g = N - 2 ;~~~~~ dim(G)= {(N^2 - N)\over 2}}
     
As discussed in \gvconjecture\ the existence of gravitational
anomaly should be accompanied by terms from topological
string amplitudes which
contain the characteristic class
$R\wedge R$.  These terms can only enter if constant maps
contribute.
The relevant topological
string  in the present context 
is the one on the resolved conifold geometry modded out by
the orientifold action.  Let us write the anomaly term
above in terms of the parameters $t$ and $g_s$ using 
 $ (N-1) = t/g_s$ and $ g_s=2\pi i/(k+N-2)$.  Then the coefficient
of the anomaly term becomes 

 \eqn\anama{{i\pi c\over 12}={t^2(2\pi i -t)\over 48 g_s^2} + {2\pi i t
\over 48 g_s} +
 {t\over 48}}

From the above expression we notice that the sphere level term
(i.e.
O($1\over
g_s^2$ term) and torus term (i.e. $O(1)$) is exactly half of the SU(N)
answer, as should be the case because there should be no difference
between their contribution before or after orientifolding (except for an
overall factor of 1/2 as will be discussed in my detail later).  The
 term of O($1\over g_s$) should be interpreted as the contribution
from the worldsheet geometry being $RP^2$.  Since the target
is also $RP^2 $ there are no relevant ``constant maps''
(what we mean is that there is no constant map from $P^1\rightarrow P^1$
which is invariant under the $Z_2 $ action which acts on both
sides by $z\rightarrow -1/{\overline z}$).  Such a term would have
shown up at order $t^2/g_s$ which is indeed absent, as it should
(note that the order $t$ terms is somewhat ambiguous as we need
to take at least two derivatives of the $RP^2$ amplitude to fix
the symmetries).  Thus the above identification of parameters
is consistent with the expectations based on anomalies on
the dual closed string theory side.  A similar story repeats for $Sp(N)$
with $a=+1$.

\newsec{SO and Sp Chern-Simons Gauge Theories at Large N}

In \wittencs\ the partition function of a Chern Simons 
theoery with gauge group G on base manifold $S^3$ was computed as 
$S_{00}$, i.e., a particular element of the modular transformation
matrix associated to the corresponding WZW model.  On the other
hand $S_{00}$ is known for arbitrary groups
and is given by
\eqn\pfcs{Z=S_{00} = |P/Q|^{-1/2}(k+c_{g})^{-r/2}\prod_{\alpha\in\Delta^+}
2Sin({\pi (\alpha,\rho)\over( k+c_{g})}).}
Here $\alpha$ runs over the positive roots of the 
Lie algebra, $c_{g}$ is the dual Coxeter number and 
\eqn\defrh{\rho={\half}\sum_{\alpha \in \Delta^+}\alpha}
is the Weyl vector of the Lie Algebra. Here $P$ is the weight lattice and
$Q$ is the root lattice. $|P/Q|$ refers to the cardinality  of the
quotient
space and $r$ is the rank of the group.
Note that
$(\alpha,\rho)$ is an integer or half integer for every $\alpha$. 
Let $(\alpha,\rho)$ take the value $j$, $f(j)$ times 
as $\alpha$ runs over all positive roots. Then
 \eqn\pfa{Z  = |P/Q|^{-1/2}(k+c_{g})^{-r/2}2^{|\Delta^+|}\prod_{j} 
 Sin ( {\pi j\over k+c_{g}})^{f(j)}.}
Consequently, the free Energy is 
\eqn\freen{F = -log(Z) =1/2log|P/Q|+ {r\over 2}log(k+c_{g}) -
 |\Delta^+|log2- \sum_{j} f(j)\log( Sin ({\pi j\over k+c_{g} })).}

We will compute the sum in \freen\ to find an explicit formula 
for $F$ for the case of $SO(N)$ and $Sp(N)$ gauge groups.
Define
\eqn\lmdef{\lambda =-it={ 2\pi (N+a)\over(k+c_{g})}}
where $a=-1$ for $SO(N)$ gauge group and $a=+1$ for $Sp(N)$.
 The crucial piece in Free energy is the last term in \freen ;
(the other terms will introduce  a minor modification 
of the final result which we will note below). Let us continue
calling this term by $F$.
In other words we write
\eqn\frew{F =  -\sum_{j} f(j)\log\left( 
Sin({\pi j \lambda \over 2\pi(N+a)})\right)}
Using the product formula  for $Sin(\pi x)$
\eqn\prodform{Sin(\pi x) = \pi x
 \prod_{n=1}^\infty (1-({x\over n} )^2)} 
\frew\ becomes
 \eqn\fn{
F = -\sum_j f(j) \sum_{p=1}^{\infty} \ln(1 - { j^2\lambda^{2}\over
 4(N+a)^{2}p^{2}\pi^{2}}) - \sum_j f(j)log({j\lambda \over
2(N+a)})}

The last term in \fn\ again is simple
and we will incorporate it in the computation at the end.
Let us concentrate on the first term, and still call it by
$F$.
On using  the expansion
\eqn\logexp{log(1-a) = -\sum_{m=1}^\infty {a^{m}\over m}}
first term in \fn\ becomes
\eqn\prfr{
\sum_{m=1}^{\infty} \sum_j f(j) 
({j\lambda \over 2\pi (N+a)})^{2m}  { \zeta(2m)\over m}.} 
In order to evaluate $F$ we must thus evaluate the sum 
\eqn\suma{\sum_j f(j)j^{2m},} 
which we will now turn to.

\subsec{$SO(N)$ with $N$ Even}
The computation depends on which group
we are dealing with and it is simplest for the simply
laced group $SO(N)$ with $N$ even, which we will first compute.
In order to evaluate $F$ we must thus evaluate the sum 
\eqn\suma{\sum_{j=1}^{N-2} f(j)j^{2m}.}

$f(j)$  may be computed by considerations of the
root lattice of $SO(N)$ with the result  
\eqn\fjcomp{f(j)= \cases {
                       {(N+1-j) \over 2} & $j$ odd $< N/2$ \cr
                       {(N-1-j)\over 2} &  $j$ odd $\geq N/2$ \cr
                         {(N-j)\over 2} & $j$ even $< N/2$ \cr
                         {(N-j-2) \over 2} & $j$ even $\geq  N/2$.}}
Thus the summation \suma\ reduces to 
\eqn\nsuma{\sum_{j=1}^{N-2} {(N-1-j)\over2} j^{2m} + 
 \sum_{j=1}^{(N/2 - 1)}j^{2m} 
   -2^{2m-1} \sum_{j=1}^{(N/2 - 1)} j^{2m}.}

Consequently, \prfr\ takes the form
\eqn\freec{ \eqalign{
F=& \sum_{m=1}^{\infty} \sum_{j=1}^{N-2} {(N-1-j)\over 2}j^{2m}
({\lambda \over
2\pi(N-1)})^{2m} {\zeta(2m) \over m} \cr 
& + \sum_{m=1}^{\infty}(1- 2^{2m-1})  
\sum_{j=1}^{{N\over2} -1} j^{2m}( {\lambda\over 
2\pi(N-1)})^{2m} {\zeta(2m)\over m} .\cr }}
Performing the summation over $j$ using the formulae
$$\sum_{j=1}^k j^l={(k+{1\over 2})^{l+1}\over l+1}
+\sum_{g=1}^{[{l\over 2}]}{2^{1-2g}\over (l+1)} {{l+1}\choose{2g}}(-1)^{g-1}B_g(1-2^{2g-1})
(k+{1\over 2})^{l+1-2g}$$
$$\sum_{j=1}^k j^l={(k+1)^{l+1}\over l+1}-{1\over 2}(k+1)^l
+{1\over l+1} \sum_{g=1}^{[{l\over 2}]} {{l+1}\choose{2g}}
 (-1)^{g-1}B_g (k+1)^{l+1-2g}$$
we find that 
 the first term in the \freec\ is exactly half of the $SU(N)$ answer and
is given by the first three terms below and the second term above
leads to the last two terms below:
\eqn\freed{{\eqalign{
F=& \sum_{m=1}^{\infty}(N-1)^2{\zeta(2m) \over 2m(2m+1)(2m+2)}({\lambda
\over 2\pi})^{2m} - \cr 
&\sum_{m=1}^{\infty}{\zeta(2m)\over 4m}({\lambda \over 2\pi})^{2m} +\cr
& \sum_{g=2}^{\infty}(N-1)^{2-2g}{(-1)^gB_g \over
4g(2g-2)}\sum_{m=1}^{\infty}\zeta(2g-2+2m){2g-3+2m \choose 2m} ({\lambda
\over 2\pi})^{2g-2+2m} + \cr 
&\sum_{m=1}^{\infty}(N-1) ({ 1-2^{2m-1} \over 2^{2m} })
{\zeta(2m) \over 2m(2m+1)} ({\lambda \over 2\pi})^{2m} + \cr  
&\sum_{m=1}^{\infty}  
2(N-1)^{1-2g} ( 1-2^{2m-1} )( {\lambda \over 2\pi })^{2m} 
( {1\over2} )^{2m+1-2g}  {\zeta(2m) \over 2m(2m+1) } {2m+1 \choose 2g}  
(-1)^{g-1} B_{g} ( 2^{1-2g}-1 ).\cr}}}

The first three terms in the above expression correspond to half of
the SU(N) answer and is given by \zeroandone\ and \twoorlarger\ . We
rewrite those results here with the correct factor of half. 
So the first term in
\freed\ gives the tree level answer which is :

\eqn\tree{
\eqalign{ F_0 & = {-\zeta(3)\over 2} + {i\pi^2\over 12} t -i\left( {m\over 2} + {1\over 8}\right)
\pi t^2 + {i \over 24}t^3 + \sum_{n=1}^\infty {n^{-3}\over 2} e^{-nt} \cr}}

The second term in \freed\ gives the one loop answer and is given by :
\eqn\oneloop{F_1  = {1 \over 48} t + {1 \over 24} 
\log \left( 1-e^{-t} \right) }

The third term in \freed\ gives the higher loop answer and is given by :
\eqn\higherloop{F_g = {(-1)^{g-1} \over 4g(2g-2)} B_g \left[ 
{(-1)^{g-1} \over (2\pi)^{2g-2}} 2\zeta(2g-2)
        - {1 \over (2g-3)!} \sum_{n=1}^\infty n^{2g-3} e^{-nt}
\right].}

The fourth term corresponding to $O(1/g_s)=O(N)$ term
(which will correspond to the worldsheet $RP^2$ contribution is: 
\eqn\freek{(N-1)
\sum_{m=1}^{\infty} ({ 1-2^{2m-1} \over 2^{2m} })
{\zeta(2m) \over 2m(2m+1)} ({\lambda \over 2\pi})^{2m}}
\eqn\freel
{{(N-1)2\pi \over \lambda} \sum_{m=1}^{\infty}\left({2\over 2^{2m+1}} -
{1\over 2}\right){\zeta(2m)\over 2m(2m+1)}\left(\lambda\over
2\pi\right)^{2m+1} }

 Writing the above expression in terms of $g_{s}$ and $\lambda$ 
\eqn\freem
{{2\pi i \over g_{s}}\sum_{m=1}^{\infty}\left({2\zeta(2m)\over
2m(2m+1)}\left({\lambda \over 4\pi}\right)^{2m+1} - {\zeta(2m)\over
4m(2m+1)}\left(\lambda \over 2\pi\right)^{2m+1}\right) }
The above expression can be rewritten as:
 \eqn\freen
{ { 1\over g_{s}}{\left\{ \sum_{n= odd 1,3..}^{\infty}{1\over
n^{2}}\exp(-nt/2) + at+b\right\}}} 
The above term also includes the contribution
of the terms we dropped before which only enter into
the terms $at +b$.

Finally, the last term in \freed , upon
substituting $m=g+p-1$, takes the form
$$\sum_{g=1}^{\infty} 2(N-1)^{1-2g}{(-1)^{g-1}B_{g} \over
2g(2g-1)}(1-2^{2g-1})$$
\eqn\freef{\sum_{p=2-g}^{\infty}( ({\lambda \over 4\pi})^{2(g+p-1)}\zeta (2g+2p-2)
{2g+2p-3 \choose 2g-2} -{ 1\over 2}({ \lambda \over 2\pi})^{2(g+p-1)}\zeta
(2g+2p-2) {2g+2p-3 \choose 2g-2})}
 After doing the summation over $p$, the $\lambda$ dependent part in
\freef\
becomes 
\eqn\freeg{\sum_{g=1}^{\infty}({\lambda \over N-1})^{2g-1}{(-1)^{g-1}B_{g} 
\over
4g(2g-1)}(1-2^{2g-1})\left\{\sum_{n \in odd Z} {1 \over (2\pi n
+\lambda)^{2g-1}}  - \sum_{n \in even Z}{1 \over (2\pi n + \lambda)^{2g-1}}
\right\} }
This, when written in terms of string variables i.e. $g_{s}$ and Kahler
parameter $t$ the above expression becomes
\eqn\freeh{\sum_{g=1}^{\infty}(g_{s})^{2g-1}{(-1)^{g-1}B_{g} \over
4g(2g-1)}(1-2^{2g-1})\left\{\sum_{n \in odd Z} {1 \over (2\pi i n 
+t)^{2g-1}}  - \sum_{n \in even Z}{1 \over (2\pi i n + t)^{2g-1}}
\right\} }
 In terms of the worldsheet instantons (i.e.$\exp{-t}$)
  this expression becomes 
\eqn\freej{\sum_{g=1}^{\infty}(g_{s})^{2g-1}{(-1)^{g-1}B_{g}
\over (2g)!}{\left(
1- {1\over 2^{2g-1}}\right)} \sum_{k = odd 1,3..}^{\infty}
k^{2g-2}\exp(-kt/2) }

In terms of the worldsheet instanton i.e. Kahler parameter $t$ and
 $\exp(-t)$ this becomes, including the $O(1/g_s)$ term,
\eqn\freez{{1\over 2}\left\{
\sum_{n=1}^{\infty}\left({\exp(-nt/2)\over 2n Sin(ng_s/2)}\right)
-\sum_{n=1}^{\infty}(-1)^{n}\left({\exp(-nt/2)\over 2n
 Sin(ng_s/2)}\right)\right\}+{at+b\over g_s}+c}
where $a,b,c$ include the contribution
of the terms we neglected above.

\subsec{SO(N) theory when N is odd}

  We do the calculation for SO(N) where N is odd; 
Here also we define $c_g$ and $\lambda$ in the same way as in
the previous case i.e. 
\eqn\deef{c_g = N-2;~~~~~~~~ \lambda = {2\pi (N-1)\over (k+N-2)}}
\eqn\codef{g_{s} = {2 \pi i \over (k+N- 2)} ;~~~~~~ 
t = i \lambda }
$f(j)$ in this case differs from the previous case and is given by
\eqn\fjcomu{f(j)= \cases {
                         1  &   $j = {2k-1 \over 2}$  ($k$= 1,..., ${(N-1) \over
                           2})$  \cr 
                          {(N-1-j) \over 2} & $j$ even \cr
                          {(N-2-j) \over 2} & $j$ odd }}                  
Thus the summation \suma\ reduces to 
\eqn\nnsuma{\sum_{j=1}^{N-2}{(N-1-j)\over 2}j^{2m} + {(1- 2^{2m-1})\over
2^{2m}}\sum_{j=odd 1,3..}^{N-2}j^{2m} }

So carrying out the summation in the expression
\eqn\freeo{\sum_{j}f(j)j^{2m}{\left({\lambda\over2\pi(N-1)}\right)^{2m}{\zeta(
2m)\over m} }}

We first sum over $j$ and then substitute $m = g+p-1$ and then sum over $p$
to get exactly the same expression as was obtained in $SO(N)$ where N is
even,
as is expected for a consistent large $N$ analysis of $SO(N)$ gauge theory,
which should not be sensitive to the parity of $N$.

\subsec{Sp(N) theory}

 We use the notations in which N is even.
\eqn\valgp{c_{g}={N\over 2}+1;~~~~~~~~~ \lambda = {2\pi (N+1) \over (k+{N \over 2}+1)}}
and 
\eqn\codef{g_{s} = {\pi i \over(k+{N\over 2}+ 1)} ;~~~~ t = i {\lambda
\over 2}}
$f(j)$ in this case is given by the following expression 
\eqn\fjcomun{f(j)= \cases {
                         {(N-1-j) \over 2}    &   $j$ odd $\leq {N \over 2}$
                             \cr 
                          {(N-j) \over 2} & $j$ even $< {N \over 2} $\cr
                          {(N+1-j) \over 2} & $j$ odd $>{N \over 2}$ \cr
                           {(N+2-j)\over 2} & $j$ even $>{N \over 2}$     }} 

Thus the summation \suma\ reduces to 
\eqn\freew{\sum_{j=1}^{N}\left({(N+1-j) \over 2} \right)j^{2m} + 
\left({2^{2m-1}-1}\right)\sum_{j=1}^{N/2}j^{2m}}
Comparing this to \nsuma\ we see that except for the sign change
on the second term, and replacing $N-1$ by $N+1$, it is exactly
the same result, and we can thus readily write the result,
which is summarized below.

\subsec{Summary}

Let us now summarize what we have found.  For the case of $SO(N)$
we have found that the partition function of the Chern-Simons
theory on $S^3$ can be written in terms of the natural variables
of the closed topological string $(g_s,t)=({2\pi i\over k+N-2},i{N-1\over
k+N-2})$
 as
\eqn\finex{F_{SO}(g_s,t)={1\over 2} \sum_{n=1}^{\infty} {e^{-nt}\over n [2\ {\rm 
sin}(n g_s/
2)]^2}+{1\over 2}\left\{\sum_{n=1}^{\infty}
\left({{\rm e}^{-nt/2}\over 2n \ {\rm sin}(ng_s/2)}\right)
-\sum_{n=1}^{\infty}(-1)^{n}\left({{\rm e}^{-nt/2}\over 
2n \ {\rm sin}(ng_s/2)}\right)\right\}}
where there is in addition a finite polynomial of order three in $t$.
The $Sp$ answer is similar to the $SO$ case and is given
in terms of $(g_s,t)=({2\pi i\over k+{N\over 2}+1},i{N+1\over
k+{N\over 2}+1})$ as
\eqn\finsp{F_{Sp}(g_s,t)={1\over 2} \sum_{n=1}^{\infty} {e^{-nt}\over n [2\ {\rm 
sin}(n g_s/
2)]^2}-{1\over 2}\left\{\sum_{n=1}^{\infty}
\left({{\rm e}^{-nt/2}\over 2n \ {\rm sin}(ng_s/2)}\right)
-\sum_{n=1}^{\infty}(-1)^{n}\left({{\rm e}^{-nt/2}\over 
2n \ {\rm sin}(ng_s/2)}\right)\right\}}
again up to a finite polynomial in $t$. Note that
\eqn\relat{F_{Sp}(g_s,t)=F_{SO}(g_s,t+{2\pi i}) }

\newsec{Dual Topological string Interpretation}

Now we try to interpret these results in the context of topological strings
on the resolved conifold, modded out by a $Z_2$ orientifold.
Namely we consider the $O(-1)+O(-1)$ geometry over ${\bf P}^1$
modded out by an antiholomoprhic involution, which in a local
coordinate chart looks as $(z_1,z_4,z)\rightarrow ({\overline z_4},
{\overline z_1},{-1\over {\overline z}})$, where $z_1,z_4$
are coordinates along the fiber and $z$ is the coordinate along ${\bf P}^1$.
On the worldsheet theory, we consider all closed orientable
and non-orientable Riemann surfaces.  It is well known that the
non-orientable ones can be obtained from orientable Riemann surfaces
by including one or two ``crosscaps'', where a crosscap corresponds
to a disc removed from the oriented Riemann surface and where the boundary
points of the disc
are identified by a 
reflection\foot{Three crosscaps can be traded for a single crosscap
and a handle.}.  The notion of orientifolding means that
we consider maps from the Riemann surfaces to the target
for which the worldsheet
$Z_2$ involution at crosscaps are compatible with the anti-holomorphic
$Z_2$ involution in the target.  The Euler characteristic of a
closed Riemann surface of genus $g$ is $\chi_0 =2-2g$.  The non-orientable
ones obtained by adding one crosscap to a genus g surface has
Euler characteristic $\chi_1=1-2g$, and the non-orientable ones
obtained by adding two crosscaps have $\chi_2=-2g$.
The difference between the $SO$ theories and the $Sp$ theories
is that the Riemann surfaces with an odd number of crosscaps
have a different relative minus sign.  Note that 
the string partition function is weighted with $g_s^{-\chi}$ 
and so we see that for a non-orientable Riemann surface
with a single crosscap we have only {\it odd} powers
of $g_s$, whereas for even number of crosscaps the
power of $g_s$ is even.

Now we are ready to analyze the predictions \finex\ and \finsp\
for closed topological string amplitudes. Note that the first
term in both of them is given by half of the $SU(N)$ answer.
Since the $SU(N)$ answer is given by orientable Riemann surfaces,
and that is also part of what we should sum over here, that is as
expected.  We can also explain the overall factor of $1/2$:  When
we mod out by a symmetry of order $|G|$ the genus $g$ amplitudes
of closed Riemann surfaces will get an extra weight of $1/|G|^g$
(the Hilbert space interpretation is the projection operator
acting for each handle). Here $|G|=2$.
 Since the genus $g$ amplitude for
orientable Riemann surface as genus $g$ is weighted by
$(g^{SU}_s)^{2g-2}$, by redefining the string coupling
$g^{SO}=g^{SU}/{\sqrt 2}$ we see that we get an overall
factor of $1/2$ in front of the answer we got for
the closed oreintable string theory.  This explains the
first terms in \finex\ and \finsp .

In addition we have to consider non-orientable Riemann surfaces with one or
two crosscaps. However, the extra term in \finex\ and \finsp\ have
only odd powers of $g_s$ which implies that they correspond
to non-orientable Riemann surfaces with a single crosscap.  Thus
the partition function in this background corresponding to an even number
of crosscaps must be zero.  This is remarkably consistent with the fact
that the terms with odd power of $g_s$ differ just by an overall sign
between the $SO$ \finex\ and the $Sp$ \finsp\ 
cases, as is expected for Riemann surfaces
with an odd number of crosscaps.

Let us consider the amplitude corresponding to $RP^2$ (i.e. $g=0$
with one crosscap).  This corresponds to keeping the term
of order $1/g_s$ in \finex\ (which up to
an overall sign is the same as that for \finsp ):
 \eqn\frene
{ { 1\over g_{s}}{\left\{ \sum_{n= odd 1,3..}^{\infty}{1\over
n^{2}}\exp(-nt/2) + at+b]\right\}}.}
The exponential terms
 should correspond to holomorphic maps from ${\bf P}^1$
to ${\bf P}^1$ which are invariant under the simultaneous
operation of $z\rightarrow -1/{\overline z}$ on both ${\bf P}^1$'s.
It is easy to see that this corresponds to the identity map and its
odd covers (for example the $2n$ order cover $z(z')=(z')^{2n}$ or any
other even order cover
is not compatible with the $Z_2$ actions).
This is a strong check for our proposed conjecture.

\subsec{Schwinger Interpretation}
We can try to check the predictions for the topological string amplitudes
by connecting the predictions of topological strings to superpotential
terms in superstring propagation on the
corresponding CY, as was done in
 \gopva\ for the orientable case.
 The case at hand is similar to \ov\ and we will consider a similar
embedding for this purpose.  Consider type IIA superstring on the
resolved conifold geometry $O(-1)+O(-1)$ over ${\bf P}^1$ times
$R^4$.  Mod out by the orientifold action we have discussed for
the internal Calabi-Yau, which also acts as $(-1,-1,1,1)$ 
on $R^4$.  This gives a theory in $1+1$ dimensions (the invariant
directions for the orientifold action in $R^4$), with 4 supercharges.
This is similar to \ov\ except that there, instead of orientifolding
one put some $D4$ branes in the resolved conifold geometry whose
worldvolume consists of a Lagrangian submanifold of the Calabi-Yau
times an $R^2$ subspace of Minkowski space. Thus the same arguments
as \ov\ applies to this case.  In particular turning on
graviphoton field strength, relates the topological
string amplitudes to Schwinger-like one loop computations
of particles coupled to a background of constant field strength.
The particles being related to wrapped D2 branes (with some
number of D0 branes bound to them).
The steps are exactly as in \ov\ and so we will not repeat them here.

The three different terms in \finex\ can now be interpreted
accordingly.  The first term correponds to a D2 brane wrapped
over ${\bf P}^1$, of BPS mass $t$, and propagating in $R^4$ (which explains
in particular the fact that there are two powers of ${\rm sin}$ 
in the denominator).  The next two terms should correspond
to particles moving in 2-dimensions, as there is only a single
power of ${\rm sin}$ in the denominator. Moreover its BPS mass is $t/2$.
This can be easily understood:  Due to orientifolding the geometry
if we consider a single $D2$ brane, whose
worldline passes through the fixed point of $Z_2$ action
in $R^4$ which wraps only half of the ${\bf P}^1$ whose
boundaries are identified due to the $Z_2$ action, it gives rise
to a particle in 2 dimensions, with BPS mass $t/2$. The fact
that there are two terms of this type in \finex\ is also easily
understood. We can consider putting a single $D0$ brane
dissolved in the $D2$ brane.  Due to the $Z_2$ action
this counts as a fractional brane with 1/2 units of $D0$ brane charge.
In other words the existence of two terms reflects the fact that
the unit of $D0$ brane charge has changed due to the $Z_2$ orientifold
action.  Notice that one differs from the other (up
to an overall sign) by shifting of $t\rightarrow t+
2\pi i$, which is what is expected for the effect of an extra $D0$ brane.
The explanation of the relative sign between the two terms as well as
between
\finsp\ and \finex\  must be due to changing the fermion number assignment for
these particles.

\newsec{Connection With ${\cal N}=1$ Systems in $D=4$}

We can embed this duality into type IIa superstring and
deduce some duality involving an ${\cal N}=1$ gauge system
in four dimensions.  For the case of $SU(N)$ this was done
in \vaau .  This was also recently
reinterpreted as a geometric duality by embedding of type IIa
strings in M-theory \amv .  The situation at hand involves
modding out by an extra $Z_2$ operation on both sides, and so
it should go through.

Let us discuss briefly how this works:  Consider type IIa
strings in the deformed conifold background $T^* S^3$
with $N$ units of $D6$ branes wrapped on $S^3$.  We mod
this out by an orientifold $Z_2$ which in the internal Calabi-Yau
preserves the $S^3$ and acts trivially on $R^4$.  This gives
rise to $SO(N)$ or $Sp(N)$ gauge group living
on the brane depending on the choice
of the sign for the crosscap.  In particular this leads to
a sector of the theory with ${\cal N}=1$ supersymmetric
$SO(N)$ or $Sp(N)$ Yang-Mills theory.  

In \vaau , which corresponds to the $SU(N)$ Yang-Mills
theory, the moduli field
associated to the (complexified) blow up mode ${\bf P}^1$ was identified as the
vev of gaugino bilinear $t=S=g_sTr {\cal W}^2$.  The lowest component
of the field gets a vev which means that we have
$\langle Tr \lambda^2 \rangle \not= 0$.  The value of the
modulus was determined by extremizing a superpotential $W$
which depends on $N$, and the bare coupling of the gauge theory.
In the limit that the bare coupling is small we have a decoupled
gauge theory system and the superpotential becomes essentially
$$W=N S {\rm log} S+a S$$
and $dW/dS =0$ gives $S^N={\rm const}$ which is in agreement
with the $N$ vacua expected for the ${\cal N}=1$ Yang-Mills
theory.  The term proportional to $N$ above, comes from
the fact that there are $N$ units of $RR$ flux piercing
through the $S^2$.

In the case at hand we are acting by orientifolding
on both the gauge theory side and its dual.  On the dual
side we have no branes left and we have the type IIa background
on the orientifold of the blownup resolution of the conifold
times Minkowski space, where the orientifold acts in the
internal part as described before and acts trivially on the
Minkowski spacetime.  The story is similar to that in \vaau\
with two  minor differences:  First 
of all the flux is halved by the orientifolding
operation and also shifted by the fact that on the gauge  theory side
we also have an orientifold which does carry RR-charge of $\mp 4$ units,
for $SO,Sp$ cases repsectively which leads to
${N\over 2}\rightarrow {N\over 2} \mp 2$.  In addition we will now also
have a superpotential coming from the ${\rm RP}^2$ worldsheet
(the term proportional to $1/g_s$).  In the small 
$t=S$ limit this give $\pm  S {\rm log} S$ and so altogether
in the decoupled limit we have
$$W=({N\over 2} \mp 1) S{\rm log}S +a S$$
which gives ${N\over 2} \mp 1$ vacua.  This agrees with the expected
answer for the $O(N)$ and $Sp(N)$ case \ref\wittvec{E. Witten,
``Toroidal Compactifications Without Vector Structure,''
JHEP 9802 (1998) 006.}.
  Also whether there are $N\mp 2$ vacua
or half as many depends on the normalization assigned to $W$
and may reflect the ambiguities in the global choices for
groups \wittvec\ which would be interesting to better understand in connection
with the global issues in realization of gauge group
in string theory.

\subsec{Embedding in M-theory}

One can also follow \amv \ref\ach{B.S.
Acharya,``On Realising
${\cal N}=1$ Yang-Mills in M-theory,'' hep-th/0011089.}\
and embed this construction in M-theory
in the context of a $G_2$ holonomy manifold which is topologically
$R^4\times S^3$ modded out by a discrete group.
This is easiest to do for the case of $SO(even)$ which 
is the only case we will consider here\foot{The non-simply
laced case can be obtained by introducing a
suitable $Z_2$ involution.}.
The only modification compared to \amv\ in deriving
the large $N$ duality for type IIa strings is that there is an extra
$Z_2$ action before and after the $S^3$ flop.  
We write the 7-fold as
$$|z_1|^2+|z_2|^2-|z_3|^2-|z_4|^2=V$$
where the $S^3$ before the flop ($V>0$) is identified
with the locus $z_3=z_4=0$.
Mod out by the Dihedral group generated by
$$(z_3,z_4)\rightarrow (\omega z_3 ,\omega^{-1} z_4)$$
$$(z_3,z_4)\rightarrow (z_4,-z_3)$$
where $\omega^{{N-4}}=1$.  We are assuming
$N$ is even.  
Note that by introducing a complex conjugate
variable for $z_4\rightarrow {\overline z}_4$, this action can
also be viewed as
$$(z_3,z_4)\rightarrow (\omega z_3,\omega z_4 )$$
$$(z_3,z_4)\rightarrow ({\overline z_4}, -{\overline z_3})$$
This way of writing it is easier to use in
inferring its dual description.

The fixed locus of this action
is $S^3$ and it has a D-singularity corresponding
to $SO(N)$ gauge symmetry.  The flopped geometry
gives the dual theory.  This is where $V<0$.  In this
case the group action has no fixed points. 

To interpret this in terms of type IIa string we have to
choose the `11-th' direction.
We choose it to be
 identified with the circle
$(z_3,z_4)\rightarrow (e^{i\theta}z_3,e^{i\theta}z_4)$.
In this identification it is easy to see that with $V>0$
we have $N$ D6 branes, which have been orientifolded
with $S^3$ times the Minkowski space, being identified
as the orientifold plane.  For $V<0$ we have the
same fibration giving rise to a Hopf fibration $S^3\rightarrow S^2$
where the complex coordinate on $S^2$ is identified with $z=z_3/z_4$.
Moreover the fact that the group we are modding out has a cyclic
element of order $N-4$ in the direction of the eleventh circle, 
implies that in type IIa perspective we have $N-4$ units of
RR flux through the $S^2$ (note that it does not act on $z$).
The extra $Z_2$ generator acts by taking $z\rightarrow -1/{\overline z}$
(and acting also in some way over the fiber), which we
identify with the orientifold action we have discussed 
above in the context of topological
 strings.  The
identification of parameters in the M-theory, ignoring Euclidean
M2 brane instantons leads, as in \amv\ to the formula $t=-V/(N-4)g_s$ where
$V$ is the volume of $S^3$ before transition and $t$ is the size
of the $P^1$, all in type IIa string units.  This would naively
suggest $N-4$ vacua. This in fact would agree with the naive formula
one would get from Type IIa  string perspective if
one ignores worldsheet instantons.
 However as discussed above for small $t$ the extra
superpotential term, coming from the $RP^2$ worldsheet geometry,
which lifts up to Euclidean M2 brane instantons in M-theory,
will shift this to $N-2$ vacua.

\vglue 1cm

We would also like to thank Freddy Cachazo, Rajesh Gopakumar and Sheldon Katz
for valuable discussions.

\vglue 1cm 
 This research is supported in part by NSF grants PHY-9802709
and DMS 9709694.

\listrefs
\end